\documentclass[useAMS,usenatbib]{mn2e}

\title[An accretion disk and radio spectra of pulsars]{An accretion disk and radio spectra of pulsars}
\author[F.V.Prigara]{F.V.Prigara\thanks{E-mail:fprigara@imras.yar.ru}\\
Institute of Microelectronics and Informatics, Russian Academy of
Sciences, 21 Universitetskaya, Yaroslavl 150007,Russia}

\begin{document}

\pagerange{\pageref{firstpage}--\pageref{lastpage}} \pubyear{0000}

\maketitle

\label{firstpage}

\begin{abstract}
On the basis of the unified model of compact radio sources, the
dependence of a turnover frequency in the smoothed radio spectrum
of a pulsar upon the ratio of the dispersion measure D to the
period P of a pulsar is obtained. This relation is produced by the
radial density wave in the accretion disk surrounding a pulsar.
The unified model of compact radio sources gives also the smoothed
spectral indices of radio emission from pulsars as $\alpha=2$ for
the gaseous disk with the temperature profile T=const and
$\alpha=3$ for the gaseous disk with the pressure profile P=const
($F_{\nu} \propto \nu ^{ - \alpha} $). The transverse density wave
in the magnetosphere of a pulsar can be responsible for the
polarisation of optical radiation from pulsars.
\end{abstract}


\section{Introduction}

The mechanism of radio emission from pulsars has remained so far
an unsolved problem \citep{b5,b20}. The difficulties which have
been encountered by plasma mechanisms of radio emission from
pulsars clearly show that the radiation is produced by a
low-energy medium \citep{b8}. Such a medium is the gaseous disk
surrounding the central star.

The accretion disk surrounding a radio pulsar is fed by the
fallback material left over after the original supernova explosion
\citep{b3}. Note that ordinary, not recycled pulsars normally are
not binaries \citep{b5}. Pulsar wind \citep{b7} may be responsible
both for a stationary convection in the accretion disk surrounding
a pulsar and for outflows of gas in the gaseous disk.

This approach allows us to consider radio pulsars as
representatives of a wide class of compact radio sources which
includes also active galactic nuclei and maser complexes
\citep{b16}. The unified model of compact radio sources, as it is
shown below, gives such characteristics of the smoothed radio
spectra of pulsars as the spectral indices, a short-wavelengths
cut-off and turnover frequencies.

In the unified model of compact radio sources, radio emission from
pulsars is treated as thermal radiation from an accretion disk
amplified by the maser mechanism \citep{b16}. A maser
amplification of thermal radio emission in continuum produces the
high brightness temperatures of compact radio sources and a rapid
variability of total and polarised flux density, that is
characteristic for non-saturated maser sources. In particular,
pulsars signals show a variability on every observable timescale
up to nanoseconds \citep{b5,b20}.

\section{Gaseous disk model}

It was shown recently \citep{b16} that thermal radio emission has
a stimulated character. According to this conception thermal radio
emission from non-uniform gas is produced by an ensemble of
individual emitters. Each of these emitters is a molecular
resonator the size of which has an order of magnitude of mean free
path \textit{l} of photons

\begin{equation}
\label{eq1}
l = \frac{{1}}{{n\sigma} }
\end{equation}

\noindent
where \textit{n} is the number density of particles and $\sigma $ is the
absorption cross-section.

The emission of each molecular resonator is coherent, with the wavelength

\begin{equation}
\label{eq2}
\lambda = l,
\end{equation}

\noindent
and thermal radio emission of gaseous layer is incoherent sum of radiation
produced by individual emitters.

The condition (\ref{eq2}) implies that the radiation with the wavelength $\lambda $
is produced by the gaseous layer with the definite number density of
particles \textit{n} .

The condition (\ref{eq2}) is consistent with the experimental
results by Looney and Brown on the excitation of plasma waves by
electron beam \citep{b1,b4}. The wavelength of standing wave with
the Langmuir frequency of oscillations depends on the density as
predicted by equation (\ref{eq1}). The discrete spectrum of
oscillations is produced by the non-uniformity of plasma and the
readjustment of the wavelength to the length of resonator. From
the results of experiment by Looney and Brown the absorption
cross-section for plasma can be evaluated.

The product of the wavelength by density is weakly increasing with the
increase of density. This may imply the weak dependence of the size of
elementary resonator in terms of the wavelength upon the density or,
equivalently, wavelength.

In the gaseous disk model, describing radio emitting gas nebulae
\citep{b16}, the number density of particles decreases
reciprocally with respect to the distance \textit{r} from the
energy centre

\begin{equation}
\label{eq3}
n \propto r^{ - 1}.
\end{equation}

Together with the condition for emission (\ref{eq2}) the last equation leads to the
wavelength dependence of radio source size:

\begin{equation}
\label{eq4}
r_{\lambda}  \propto \lambda .
\end{equation}

The relation (\ref{eq4}) is indeed observed for sufficiently
extended radio sources. For example, the size of radio core of
galaxy M31 is 3.5 arcmin at the frequency 408 MHz and 1 arcmin at
the frequency 1407 MHz \citep{b17}.

\section{Density profile of compact radio sources}

In the case of compact radio sources instead of the relationship
(\ref{eq4}) the relationship

\begin{equation}
\label{eq5}
r_{\lambda}  \propto \lambda ^{2}
\end{equation}

\noindent is observed \citep{b12,b13}. This relationship may be
explained by the effect of a gravitational field on the motion of
gas which changes the equation (\ref{eq3}) for the equation

\begin{equation}
\label{eq6}
 n \propto r^{-1/2}
\end{equation}

 The mass conservation in an outflow or inflow of gas
gives \textit{nvr=const,} where \textit{v} is the velocity of
flow. In the gravitational field of a central energy source the
energy conservation gives

\begin{equation}
\label{eq7} v = \left( {v_{0}^{2} + c^{2}r_{s} /r} \right)^{1/2}
\end{equation}

\noindent
where $r_{s} $ is the Schwarzschild radius. Therefore, at small values of
the radius the equation (6) is valid, whereas at the larger radii we obtain
the equation (\ref{eq3}).

It is well known \citep{b18} that the delay of radio pulses from
pulsars at low frequencies is proportional to $\lambda ^{2}$. This
fact is a mere consequence of Eq.(\ref{eq5}), if we only assume
the existence of the radial density wave travelling across the
radius with a constant velocity and triggering the pulse radio
emission. In this treatment the pulsars also obey the $\lambda
^{2}$ dependence of compact source size. Note that the wavelength
dependence of a pulse duration is a similar effect.

The spatial distribution of SiO, water, and OH masers (each of
which emits in its own wavelength) in the maser complexes also is
consistent with the $\lambda ^{2}$ dependence of compact source
size \citep{b2,b6}.

To summarise, extended radio sources are characterised by the relation (\ref{eq4}),
and compact radio sources obey the relation (\ref{eq5}).

\section{Radio emission from the gaseous disk}

The spectral density of flux from an extended radio source is
given by the formula

\begin{equation}
\label{eq8} F_{\nu}  =
\frac{{1}}{{a^{2}}}\int\limits_{0}^{r_{\lambda} }  {B_{\nu} }
\left( {T} \right) \times 2\pi rdr \quad ,
\end{equation}

\noindent
where \textit{a} is a distance from radio source to the detector of
radiation, and the function $B_{\nu}  \left( {T} \right)$ is given by the
Rayleigh-Jeans formula

\begin{equation}
\label{eq9} B_{\nu}  = 2kT\nu ^{2}/c^{2},
\end{equation}

\noindent where $\nu $ is the frequency of radiation, \textit{k}
is the Boltzmann's constant, and \textit{T} is the temperature.

. The extended radio sources may be divided in two classes. Type 1 radio
sources are characterised by a stationary convection in the gaseous disk
with an approximately uniform distribution of the temperature \textit{T$
\approx $const} giving the spectrum

\begin{equation}
\label{eq10} F_{\nu}  \approx const \quad .
\end{equation}

Type 2 radio sources are characterised by outflows of gas with an
approximately uniform distribution of gas pressure \textit{P=nkT$ \approx
$const}. In this case the equation (\ref{eq3}) gives

\begin{equation}
\label{eq11} T \propto r,
\end{equation}

\noindent
so the radio spectrum, according to the equation (\ref{eq7}), has the form

\begin{equation}
\label{eq12} F_{\nu}
 \propto \nu ^{ - 1}.
\end{equation}

Both classes include numerous galactic and extragalactic objects.
In particular, edge-brightened supernova remnants \citep{b9}
belong to the type 2 radio sources in accordance with the relation
(\ref{eq10}), whereas centre-brightened supernova remnants belong
to the type 1 radio sources.

The relationship between linear size and turnover frequency in
type 2 radio sources (gigahertz-peaked spectrum sources and
steep-spectrum sources) \citep{b14} is a consequence of the
wavelength dependence of radio source size. The turnover frequency
is determined by the equation $r_{\nu} = R$, where R is the radius
of a gaseous disk. The same equation determines a turnover
frequency for planetary nebulae \citep{b15,b16,b19}.

\section{The spectral index of radio emission}

The flux density is determined by Eq.(\ref{eq8}). In the case of
pulsars the relation (\ref{eq5}) is valid. Thus, a stationary
convection in the accretion disk with the temperature profile
\textit{T$ \approx $const} now produces the spectrum

\begin{equation}
\label{eq13} F_{\nu}  \propto \nu ^{ - 2}.
\end{equation}

An outflow or inflow of gas with the pressure profile \textit{P=nkT$ \approx
$const} gives rise to the temperature profile $T \propto r^{1/2}$, according
to Eq.(6). The flux density in this case is given by the formula

\begin{equation}
\label{eq14} F_{\nu}  \propto \nu ^{ - 3}.
\end{equation}

A combination of a stationary convection and outflows or inflows
gives the intermediate values of the spectral index $2 \le \alpha
\le 3$, where $F_{\nu}  \propto \nu ^{ - \alpha} $. The spectrum
(\ref{eq14}) seems to be the case for the Crab pulsar and many
other pulsars \citep{b18}. The spectrum (\ref{eq13}) has also been
detected in many pulsars \citep{b11}.

The range of frequencies in which the spectra (\ref{eq13}) or
(\ref{eq14}) are valid is confined by a short-wavelengths cut-off,
$\nu _{0} $, and a turnover frequency, $\nu _{t} $. In the range
$\nu < \nu _{t} $ the radio spectrum is flat or slightly inverted
\citep{b11}. This effect is similar to those in active galactic
nuclei (gigahertz-peaked spectrum sources and steep-spectrum
sources) \citep{b14}.

\section{A turnover frequency}

Consider now in more detail the delay of radio pulses from pulsars
at low frequencies discussed in Sec.3. Making use of Eq.
(\ref{eq5}), the time of the delay of pulses may be written in the
form

\begin{equation}
\label{eq15} \tau = r_{\lambda}  /u = \left( {r_{0} /u}
\right)\left( {\lambda ^{2}/\lambda _{0}^{2}}  \right) = \mu
P\lambda ^{2}/\left( {2\lambda _{0}^{2}}  \right),
\end{equation}

\noindent
where \textit{u} is the velocity of the radial density wave in the accretion
disk, \textit{P} is the period of a pulsar, $P = 2r_{0} /v$, $r_{0} $ is the
radius of the neutron star, \textit{v} is the velocity of the radial density
wave inside the neutron star, and $\mu = v/u$.

On the other hand, the delay of pulses is normally expressed in terms of the
dispersion measure, \textit{D}, as follows

\begin{equation}
\label{eq16} \tau = e^{2}D\lambda ^{2}/\left( {2\pi mc^{3}}
\right),
\end{equation}

\noindent where \textit{e} is the charge and \textit{m} is the
mass of electron, respectively \citep{b2}.

Comparing these relations, we obtain the expression for the frequency $\nu
_{0} = c/\lambda _{0} $ in the form

\begin{equation}
\label{eq17} \nu _{0} = \sqrt {\left( {e^{2}/\pi \mu mc}
\right)\left( {D/P} \right)} .
\end{equation}

The frequency $\nu _{0} $ may be interpreted as a
short-wavelengths cut-off in the smoothed radio spectrum of a
pulsar. To evaluate a cut-off frequency, we assume $\mu = 1$,
though this assumption is not likely to be valid in actual cases.
Then Eq. (\ref{eq17}) gives

\begin{equation}
\label{eq18} \nu _{0} = 0.1\sqrt {D/P} ,
\end{equation}

\noindent
where the dispersion measure, \textit{D}, is in units of $cm^{ - 3}pc$, the
period of a pulsar, \textit{P}, is in \textit{s}, and the turnover
frequency, $\nu _{0} $, is in \textit{GHz}.

The Crab pulsar has the dispersion measure of 56.8 $cm^{ - 3}pc$
and the period of 33.1 \textit{ms} (Lang 1974). The equation
(\ref{eq18}) gives $\nu _{0} $=4.2 \textit{GHz} and $\lambda _{0}
$=7 \textit{cm}, that seems to be a good estimation of the
short-wavelengths cut-off.

To obtain a turnover frequency, $\nu _{t} $, we should replace the
radius of the neutron star, $r_{0} $, by the radius of the
accretion disk, \textit{R}, in Eq. (\ref{eq15}). Then we find

\begin{equation}
\label{eq19} \nu _{t} = \nu _{0} \sqrt {r_{0} /R} .
\end{equation}

If we assume $R \approx 10^{3}r_{0} $, then for the Crab pulsar the turnover
frequency will be $\nu _{t} \approx 140MHz$, and, respectively, $\lambda
_{t} \approx 2m$. This result is in good agreement with the observed radio
spectrum of the pulsar.

\section{Conclusions}

The unified model of compact radio sources, applied to radio
pulsars, reproduces the observed spectral indices of radio
emission from pulsars. The unified model gives also the dependence
of a turnover frequency and short-wavelengths cut-off in smoothed
radio spectra of pulsars upon the ratio of the dispersion measure
to the period of a pulsar. This relation is produced by the radial
density wave in the gaseous accretion disk surrounding a pulsar.
Along with the radial density wave in the accretion disk, the
transverse density wave in the magnetosphere of a pulsar can
exist. The last produces the temporal profile of polarisation in
optical.

\section*{Acknowledgements}

The author is grateful to D.A.Kompaneets, Y.Y.Kovalev, V.N.Lukash,
V.S.Semenov, B.E.Stern and N.A.Tsvyk for useful discussions.

\label{lastpage}

\end{document}